# Seismicity observed, at Methoni seismogenic area, Greece, after the analysis of the recorded Earth's electric field of 21/2/2008 – 22/2/2008, at PYR, ATH and HIO monitoring sites, Greece.

Thanassoulas, P.C., B.Sc in Physics, M.Sc – Ph.D in Applied Geophysics.

Retired from the Institute for Geology and Mineral Exploration (IGME)
Geophysical Department, Athens, Greece.
e-mail: thandin@otenet.gr – website: www.earthquakeprediction.gr

**ABSTRACT**

The seismicity, which took place at the Methoni seismogenic area, in the time period of 20/2/2008 - 10/4/2008, is analyzed in terms of its location, time of occurrence and magnitude. Furthermore, it is compared to the tidal (T=14 days, T=1 day) lithospheric oscillation and to the epicentral area suggested by the analysis of the Earth's electric field registered on 21-22/2/2008 by PYR, ATH and HIO monitoring sites. Moreover, a comparison is made between the actual seismic energy released, during the same time period (20/2/2008 - 10/4/2008) in this specific seismogenic region and the suggested one by the probabilistic single seismic event suggested that could occur in the time period of 28/2 – 1/3/2008.

The overall analysis of the Methoni seismic event reveals the validity of the used physical models: of the lithospheric oscillation, of the lithospheric seismic energy release and the one of the homogeneous Earth used for the azimuthal intensity vector analysis of the preseismic electric signals (Thanassoulas, 2007).

**1. INTRODUCTION**

In an earlier work presented by Thanassoulas (2008), the Earth's electric field was analyzed in terms of earthquake prediction parameters (location, time magnitude) related to a probable future large EQ. This analysis was performed seven days after the occurrence of the Methoni large EQ (14$^{th}$ of February, 2008, Ms =6.7R) on the recordings (21/2 – 22/2/2008) of the Earth's electric field registered by ATH, PYR and HIO monitoring sites, located in Greece. The results obtained from this analysis suggested that a large EQ could take place in the near future in a specific: place, time and with an expected maximum magnitude, since these results fulfilled the various physical mechanisms which are often triggered and observed during some preseismic period of time before the occurrence of any large EQ.

In details, the epicentral area (**φ, λ**) of the probable future EQ was calculated by the analysis of the azimuthal direction of the Earth's electrical field intensity vector as: a. $36.40^0 / 21.61^0$ and b. $36.59^0 / 21.97^0$. This is demonstrated by green concentric circles in figure (1a).

Its time of occurrence (28/2 – 1/3/2008) was estimated by the analysis of the 14days and daily period tidal lithospheric oscillation. Specifically, the times (day and hours) when the lithospheric plate acquires its maximum stress load were assumed as the most probable times for the occurrence of a large EQ.

Its possible maximum magnitude Ms was estimated as: Ms = 7.4R while a magnitude of Ms = 6.0 - 6.5R was considered as the most probable actual one. This maximum Ms value was calculated by taking into account the regional seismic energy release in the specific seismogenic area and by applying the Lithospheric Seismic Energy Flow Model on it.

Detailed analysis of the used methodologies and more methodology application examples can be found in Thanassoulas (2007) and in URL: www.earthquakeprediction.gr

The obvious question which can be raised is: did this probable large EQ occur within the suggested seismic parameters (**λ, φ, t, M**) or it failed despite the promising results by the used methodologies?

This question will be speculated in details as follows.

**2. EPICENTER DETERMINATION**

The seismic data presented by the seismic National Observatory of Athens (NOA) show that an earthquake of magnitude Ms = 5.2R occurred on 28$^{th}$ of February 2008 in the Methoni EQ regional epicentral area. This EQ occurred within the set time window of 28/2 – 1/3/2008, set in accordance to the 14 days period lithospheric plate oscillation model. This specific EQ could be considered as a partial prediction success since it occurred within the specified epicentral area, within the specified time window but at the same time failed to conform to the suggested magnitude of Ms as 6.0<Ms<6.5R. This, theoretically, suggests that a large amount of seismic energy is still stored in the regional epicentral area. Most probably, this energy will be released within a larger period of time. Therefore, the time window was expanded from 20/2/2008 up to 10/4/2008. One more EQ was included in this study, the one which occurred at the end of 19$^{th}$ towards the 20$^{th}$ of February. This will be explained later on. In this period of time 15 EQs occurred, in the entire Greek area, with a magnitude of Ms>=5R. These EQs are presented in details (date, regional epicentral area, magnitude, compared to peaks of tidal (T=14days) oscillation) in the following Table -1.

**TABLE – 1**

| Date | Location | Magnitude | At +/- 1day from tidal peak |
|------|----------|-----------|------------------------------|
| 19/2 | Methoni  | 5.1R      | Y                            |
| 20/2 | Methoni | 6.5R | Y |
| 26/2 | Methoni | 5.7R | N |
| 26/2 | Methoni | 5.4R | N |
| 28/2 | Methoni | 5.2R | Y |
| 4/3 | Methoni | 5.1R | Y |
| 5/3 | Rhodos | 5.2R | N |
| 7/3 | Methoni | 5.3R | Y |
| 7/3 | Methoni | 5.1R | Y |
| 7/3 | Methoni | 5.0R | Y |
| 14/3 | Methoni | 5.2R | Y |
| 19/3 | Skyros | 5.5R | Y |
| 23/3 | Methoni | 5.3R | N |
| 28/3 | South Creta | 5.6R | Y |
| 30/3 | SW Turkey | 5.2R | Y |

The location of these EQs is presented in the following figure (1b) with blue circles.

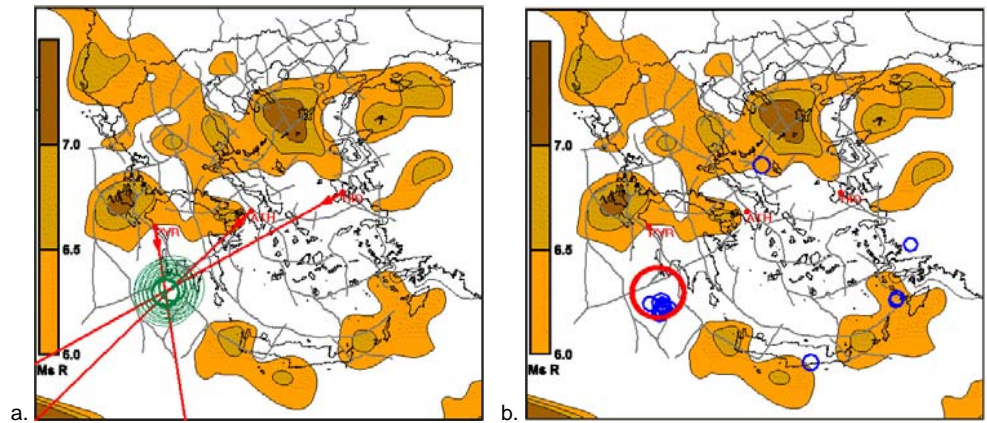

Fig. 1. Map (a) indicates, in green circles, the calculated epicentral areas which resulted from the analysis of the Earth's electric field registered at PYR, ATH and HIO monitoring sites. In map (b) the blue circles represent the EQs with Ms>=5 which occurred within the time period of 20/2 – 10/4/2008. The red circle indicates the suggested most probable regional epicentral area with a radius of 55Km.

Figure (1b) indicates that 11 EQs, out of 15, did occur within a narrow epicentral area suggested by the analysis of the Earth's electric field. The calculated success percentage (SP) for Ms>=5R is:

SP = (EQs within the epicentral area) / (total number of EQs) = 11/15 = 73.33%

**SP = 73.33%**.

This large value of the observed **SP** is attributed to facts such as: a) the main seismic event was quite large (Ms = 6.7R) and therefore aftershocks of large Ms are expected and b) post seismic electrical activity of the focal area lasted for some days more, after the occurrence of the main seismic event (14[th] of February, 2008, Ms =6.7R) thus, indicating that the regional focal area was still at a state of seismic excitement and therefore, there was unreleased much stored strain energy which could justify the generation of some more large EQs.

## 3. TIME (DAY, HOUR) DETERMINATION

The determination of the time of occurrence of a large EQ is performed in two steps. The first step is to estimate the most probable day of occurrence. This is based on the M1 tidal wave (T = 14 days) generated by the moon declination. The second step is to estimate the most probable hour (within a day) of occurrence. This estimation is performed by the use of the daily tidal waves K1 (T=24hours) and M2, S2, N2, K1 which oscillate at a period T very close to 12 hours. In both cases the oscillating lithospheric model is applied for (Thanassoulas, 2007).

In the work of Thanassoulas (2008) had been stated that the most probable day of occurrence is the time period of 28/2 – 1/3/2008 and for each specific day the daily tidal oscillation peaks had been specified as the most probable time of occurrence ( +/- 1 hour) for a future large EQ. Indeed, on 28[th] of February an EQ of Ms = 5.2R did occur in the Methoni prescribed epicentral area. This EQ deviated only for 100 minutes from the corresponding daily tidal peak of the 09h 30minutes GMT time (fig. 4).

In the next paragraph the EQs of table (1) will be studied in terms of time (day, hour) prediction.

### 3.1 Day determination.

Favorable lithospheric strain charge conditions which can trigger an earthquake are met at the peak amplitude values of the lithospheric oscillation. In terms of time prediction and specifically trying to predict the most probable day, the 14 days period tidal lithospheric oscillation (M1) is taken into account. This mode of lithospheric oscillation, for the same time period (20/2 – 10/4/2008), is presented in the following figure (2). The tidal lithospheric oscillation amplitude peaks are denoted by the letters A, B, C, D, E, F, G.

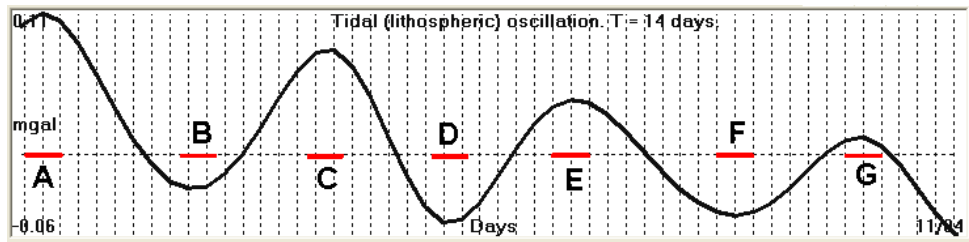

Fig. 2. Lithospheric tidal oscillation (black line, T = 14 days). The capital lettering indicates the favorable days when an EQ can be triggered. The horizontal red bars indicate the accepted time window (of +/- 1 day) from each tidal peak.

Accepting that the occurrence time-window for a future large EQ is +/- 1day from each tidal peak then the most possible days of occurrence, for any large seismic event, are the ones indicated by the horizontal red bars. Moreover, the probability (**p**) to hit "by chance" the day of occurrence of a large seismic event, in between two successive tidal peaks, is calculated as:

**p = 1/7 = 14.28%**

Therefore, it is interesting to test the validity of the tidally oscillating lithospheric model by comparing, for the same time period, the EQs which took place, against the timing of the tidal lithospheric oscillation peaks. This is presented in the following figure (3). The green bars indicate the time of occurrence of each seismic event with a lower Ms threshold of Ms = 5R.

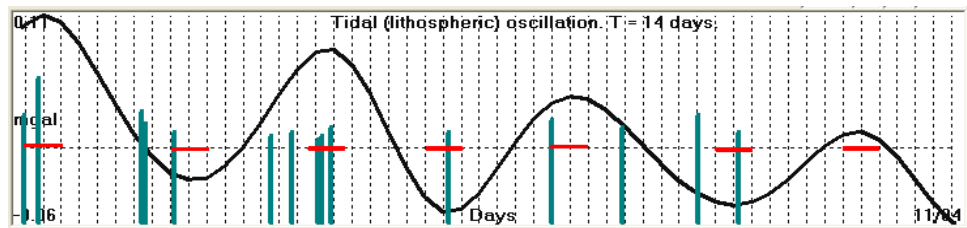

Fig. 3. EQs which occurred during the time period of 20/2 – 10/4/2008 (Ms >= 5R) are presented with green bars.

At this point it is evident why the EQ of the 19th February has been included in this analysis. This seismic event has marginally occurred at the tidal lithospheric oscillation peak of the 20th of February. At a first approach, the EQs which occurred within +/- 1 day from the corresponding tidal lithospheric peak are compared to the total number of the EQs (Ms>=5R) which occurred in the same time period in the Greek territory.

EQs which occurred within +/- 1 day from the corresponding tidal lithospheric peak     EQy = 11
EQs which occurred outside +/- 1 day from the corresponding tidal lithospheric peak    EQn = 4

**Percentage of EQs at +/- 1day from the corresponding tidal oscillation peak = 11/15 = 73.33%**

This value is far larger than the "by chance value" of **p = 14.28%** thus, it is suggested that the time of occurrence of the related seismic events is not random. Actually, it strongly depends on the mode of lithospheric tidal oscillation.
The same analysis is applied on the seismic events which did occur within the regional seismic epicentral area of the Methoni EQ.

EQs at +/- 1 day from the corresponding tidal lithospheric peak in Methoni regional epicentral area     EQ1 = 8
EQs outside +/- 1 day from the corresponding tidal lithospheric peak in Methoni regional epicentral area    EQ2 = 3

**Percentage of EQs at +/- 1day from the corresponding tidal oscillation peak = 8/11 = 72.72%**

Again the results are very similar to the previous case. It is evident that the lithospheric tidal oscillation plays a very important role in triggering the large seismic events in any seismically excited seismogenic area. The later, in return, provides the basic background required for the short-term earthquake prediction.
However, it could be argued that since a large number of large EQs followed the main event of the Methoni EQ, then the probability of having a large aftershock EQ, at the same accepted time window of a tidal peak, is large. For resolving this argument the following analysis is made.
The time window of 20/2 – 10/4/2008 equals to 50 days. In this time window there are 7 distinct tidal (M1 component) oscillation peaks. Assuming a deviation of +/- 1 day as an acceptable error from the corresponding tidal peak for the estimation of the day of occurrence, then there are 14 "favorable" days when a large aftershock EQ can occur by "chance". The "by chance" probability (CB) is calculated as:

**CB = 14 / 50 = 28%**

Even in this case, the real seismic events show a **72.72 – 73.33%** success rate which is still far larger than the "by chance" probability (CB) value.

### 3.2 Hour determination.

The time of occurrence (GMT) of the EQs of table (1) was compared to the closest daily tidal peak for their day of occurrence. An example of such a comparison is demonstrated in the following figure (4).

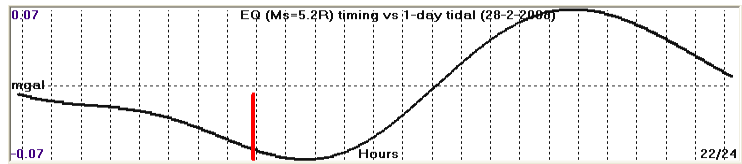

Fig. 4. Time of occurrence of a large EQ compared to the same day corresponding daily tidal oscillation. The red bar indicates the time of occurrence of the 28/2/2008 EQ. Its deviation (dt) from the closest corresponding daily tidal wave peak is: dt = 100 minutes.

All results from such comparisons for the EQs of table (1) are summarized as follows:

**Average deviation from the closest daily tidal oscillation peak = 112 minutes.**     (1)

The same comparison is made for <u>the EQs which satisfy the criterion of being near (+/- 1 day) the M1 peak.</u> In this case the average deviation from the closest daily tidal peak is:

**Average deviation from the closest daily tidal oscillation peak = 79 minutes.**     (2)

The observed better accuracy in time (hour) prediction could be explained by the fact that the two tidal components (M1, and the rest of T=12hours) of the oscillating lithospheric plate act, more or less, **in phase** upon it thus narrowing the time window in which a future large EQ can occur (in other words it increases its occurrence probability).

The mechanism of the "in phase tidal components" is intensified more for the EQs which did occur exactly on the peak of the M1 tidal wave. In such cases the time window for the occurrence of a large EQ becomes even smaller. For (3) EQs of table (1) the observed time deviation from the daily tidal peak was: dt = 30, 16 and 33 minutes. These EQs are presented in the following figures (5), (6) and (7).

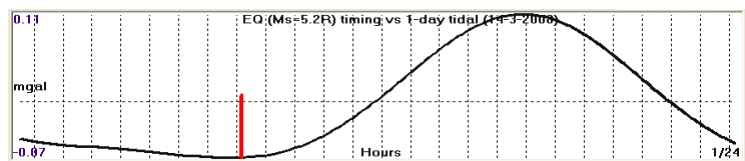

Fig. 5. Time of occurrence of a large EQ compared to the daily tidal oscillation. The red bar indicates the time of occurrence of the 14/3/2008 EQ. Its deviation (dt) from the closest corresponding daily tidal wave maximum is: dt = 30 minutes

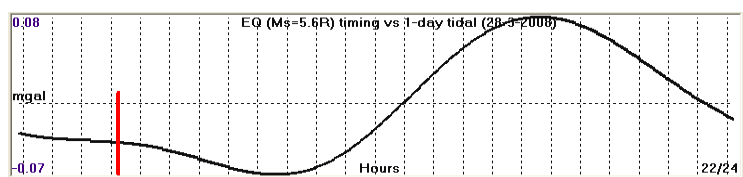

Fig. 6. Time of occurrence of a large EQ compared to the daily tidal oscillation. The red bar indicates the time of occurrence of the 28/3/2008 EQ. Its deviation (dt) from the closest corresponding daily tidal wave maximum is:  dt = 16 minutes

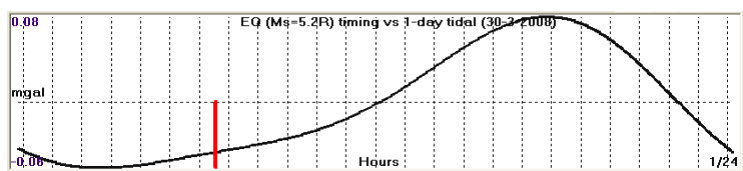

Fig. 7. Time of occurrence of a large EQ compared to the daily tidal oscillation. The red bar indicates the time of occurrence of the 30/3/2008 EQ. Its deviation (dt) from the closest corresponding daily tidal wave maximum is: dt = 33 minutes

It is evident, from this analysis, that even in an aftershocks sequence which follows a large seismic event, specific physical mechanisms (i.e tidal oscillations) still control its evolution in time, in terms of large aftershocks generation.

More similar characteristic examples (Skyros EQ, 26/7/2001, Ms=6.1R ; Lefkada EQ, 14/8/2003, Ms=6.4R ; East of Kythira EQ, 08/01/2006, Ms=6.9R) can be found in Thanassoulas (2007) and www.earthquakeprediction.gr

**4. MAGNITUDE DETERMINATION**

The third seismic parameter which was estimated by the considered study (Thanassoulas, 2008) is the magnitude of the probable future EQ. This magnitude was estimated as Ms = 6.0 – 6.5 R. However, only one EQ of Ms = 5.2R did occur in the specified epicentral area, and in the specified time window. Actually this EQ did occur on 28$^{th}$ February 2008 (see Table – 1).

The deviation of the magnitude of the EQ that did occur from the magnitude of the EQ (6.0 – 6.5R) that was probable to occur indicates that the seismic energy, stored in the regional seismogenic area, had not been totally released within the preset 3-days time window (28/2 – 1/3/2008). Consequently, this amount of energy could be released in the same seismogenic area in the near future. An inspection of the seismicity which took place within the study time period reveals that a "swarm" of EQs with Ms >= 5R did occur which created some considerable social anxiety at the same area. Actually, nine (9) EQs of Ms>=5R did occur. These EQs span from 26$^{th}$ February to 23$^{rd}$ of March 2008. After the date of 23/3/2008 no EQ of such magnitude took place in this seismogenic area until 1$^{st}$ of May/2008 when this analysis was in preparation.

Therefore, the relation of the determined stored seismic energy in the regional area before the 26$^{th}$ of February 2008 will be compared to the seismic energy released within the time period of 26$^{th}$ February – 23$^{rd}$ March 2008. This is analyzed as follows:

Each EQ from Table – 1, that did occur within the specified epicentral area, was converted in to its equivalent released energy (E). This conversion was utilized by using the following formula (1) which relates the released seismic energy (E) of an EQ to its specific magnitude (Ms). This formula has been proved valid for the Greek territory from related seismological studies (see similar formulas in any text book of seismology).

$$LogE = 11.8 + 1.5Ms \quad (1)$$

Next, all calculated energies were summed up and the resulted total energy was converted back, through the use of formula (1), into its theoretical "equivalent single EQ" magnitude Ms. The calculated theoretical magnitude Ms is:

$$Ms = 5.97R \quad (2)$$

The calculated value of Ms = 5.97R of (2) deviates for only .03R from the lower level of the range (6.0 – 6.5R) which was set as a probable magnitude for the probable case of a "single" future large seismic event. Moreover, this calculation is based only on EQs of Ms >=5R. Actually, EQs with smaller values than 5R (Ms<5R) have not been taken into account. If these EQs had be taken into account, since there is a large number of them, definitely the value of (2) would be calculated as a larger one and well within the range of Ms = 6.0 – 6.5R. This calculation will be demonstrated later on.

Therefore, the estimation of a single probable future EQ of a max magnitude of Ms = 6.0 – 6.5R failed, in the sense that instead of having the occurrence of a single large EQ, nine (9) different smaller EQs of Ms >= 5R did occur, which released almost the same amount of seismic energy.

In the following presentation, the Methoni EQ seismogenic region is studied in terms of seismic energy release. The study period of time spans from 1901 to 2008. In this study only earthquakes which did occur within the specified seismogenic area (blue line frame) have been taken into account. This is demonstrated in the following figure (8).

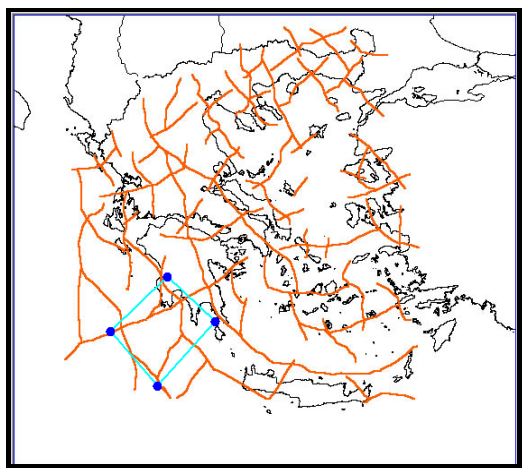

Fig. 8. Regional seismogenic Methoni EQ area is defined by the blue frame. Only EQs which did occur in this frame were taken into account.

The corresponding accumulated seismic energy release was calculated by considering all the EQs which did occur in this frame, regardless their magnitude. The graph of the accumulated seismic energy release vs. time is demonstrated in figure (9).

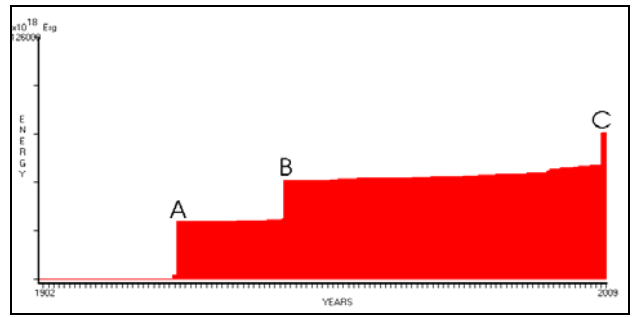

Fig. 9. Cummulative seismic energy release vs. time (1901 – 2008) calculated for the regional seismogenic Methoni EQ area. A, B, C represent, from left to right, the corresponding main large seismic events of 7/1927 (Ms = 7.1R), 10/1947 (Ms = 7.0R) and 2/2008 (Ms = 6.6R). Sampling interval of one (1) month.

It is made clear from figure (9) that the Methoni seismogenic area remains at a "lock" state for many years in between two successive large seismic events. Therefore, the "normal" cumulative seismic energy release graph (Thanassoulas, 2007) will be represented by a straight line that passes through two successive corners of this graph. An example of such a normal cumulative energy release straight line is presented in the following figure (10) and corresponds to the Methoni regional seismogenic area.

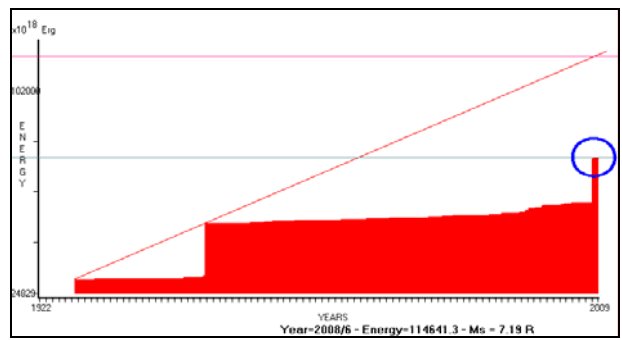

Fig. 10. The unlocked theoretical cumulative seismic energy release graph for the Methoni seismogenic area is represented by the inclined straight line. The red area of the graph represents the real cumulative seismic energy release at a sampling interval of one (1) month. Graph spans from 1922 to 2008. The blue circle area will be enlarged and analyzed in figures (12, 13).

The application of the lithospheric seismic energy flow model on these data, after the seismic event of Methoni EQ, suggests that there is still more unreleased seismic energy which could generate an EQ of max magnitude Ms = 7.19R. This calculation is straight forward from figure (10). The two horizontal lines indicate: the energy already released (lower one, after the seismic event of 14/2/2008) and what should have been released (upper one) if the seismogenic area was under normal unlock conditions. The calculated magnitude of Ms = 7.19R differs for 0.05R from the one (Ms=7.24) presented by Thanassoulas (2008). This difference is attributed to the different frames used in the previous and present studies for the Methoni seismogenic regional area. Moreover, a comparison is made between the cumulative seismic energy release calculated "before" the Methoni EQ and "after" that. To this end the next graph of figure (11) is presented.

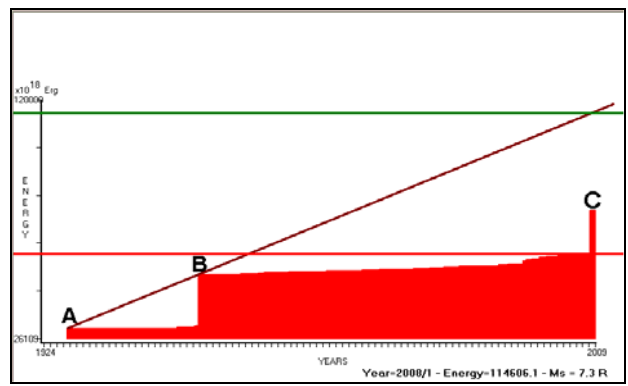

Fig. 11. The stored cumulative seismic energy just before the Methoni EQ, equals to an EQ of max magnitude Ms = 7.3 R. Sampling interval of one (1) month.

The comparison of figure (10) to the figure (11) indicates that the released seismic energy by the Methoni EQ, modified the overall stored seismic energy in the very same seismogenic area, in terms of magnitude, only for dMs = 0.11R. Actually, it conforms to the well known seismic theories that the seismogenic areas remain always at a critical state of instability. Thus, it is suggested that the Methoni seismogenic area represents still a potential seismic threat in the future.

It was referred earlier that the released seismic energy from the EQs which took place after the 20th of February, having a magnitude Ms >= 5R, equals to a single seismic event of Ms = 5.97R. The same calculation will be performed again, but for all the EQs, regardless their magnitude, which did occur in the Methoni seismogenic area for the time period of 20/2 – 01/05/2008. The later is demonstrated in the following figure (12).

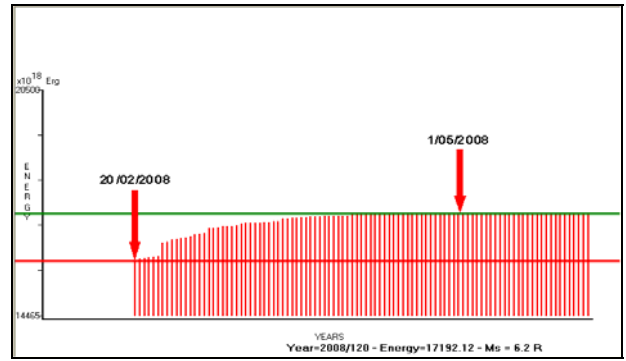

Fig. 12. Cumulative seismic energy release graph generated for the Methoni seismogenic area. The time period spans from 20/02/2008 to 01/05/2008 (red arrows). The estimated single EQ magnitude Ms is calculated as Ms = 6.2R. Sampling interval of one (1) day.

The calculated value of Ms = 6.2R conforms to the one probable single EQ with magnitude range of 6.0<Ms<6.5R suggested by Thanassoulas (2008).

What must be pointed out is that the Methoni seismogenic area, after the 20/02/2008, behaved as a "normal" one by releasing seismic energy following the "unlocked" lithospheric seismic energy flow model mode. However, after a short period of time, actually some tens of days, it was locked again, as it is shown in the following figure (13). If we consider this short period of time as a normal cumulative seismic energy release then the estimated theoretical EQ max magnitude Ms, for a virtual time of occurrence on 1/5/2008, is: max Ms = 6.23R.

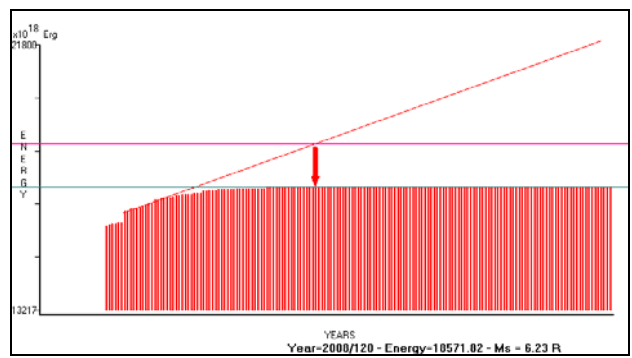

Fig. 13. Estimated theoretical EQ of max magnitude Ms = 6.23R, for a virtual time of occurrence at 1/5/2008 (red arrow). Sampling interval of one (1) day.

Consequently, the Methoni seismogenic area returned, in a short time period, to its self organized criticality mode and therefore it is possible, at any time in the future, to generate an EQ of a considerable magnitude, provided that the appropriate mechanical and triggering conditions exist.

Actually, during the fine tuning of this presentation, a large EQ (Ms = 5.6R 10/05/2008) did occur in the Methoni seismogenic area, thus verifying the previous analysis.

Generally, the seismological notion is that the end of an aftershocks sequence indicates that the stress load of the related seismogenic area has decreased back to normal levels due to the excessive seismic energy release through the aftershocks sequence itself. However, the end of an aftershock sequence can be viewed in a different way. The Methoni EQ example indicates exactly the opposite. This specific aftershocks sequence ended just because the seismogenic area "mechanically healed" and is again, in very short time period, recharged, in terms of stress load, and capable for the generation of the next large seismic event in the future.

**5. CONCLUSIONS**

The conclusion from this analysis is as follows:

The preseismic electrical signals provide quite satisfactorily the epecentral area of large EQs even if these EQs are seismic events of an aftershock sequence which was triggered by an even larger seismic event. In this respect, the epicentral area suggested by Thanassoulas (2008) could be considered as a successful one.

The time of occurrence (day, hour) of such large EQs is strongly controlled by the tidal waves which act upon the lithospheric plate and force it to oscillate accordingly. By knowing in advance the maxima and minima of its oscillation amplitude it is possible to determine the most probable time of the large EQs occurrence within a very short time period. Very small deviations of time of occurrence (less than an hour from a daily tidal peak) are generally observed when the specific EQ occurs at the M1 (T = 14 days) tidal component peak.

In the specific case, the single seismic event of the 28/02/2008 (Ms = 5.2R) satisfies the general rules set above. The "hour" estimation of the time of occurrence of this large EQ, within the time window of the 28$^{th}$ February, 2008, deviated for only 100 minutes from the time of occurrence suggested by Thanassoulas (2008).

Finally, the analysis of the released seismic energy at the Methoni seismogenic area indicated that the expected large EQ in the time period of 28/2 – 1/3/2008 of magnitude Ms = 6.0 – 6.5R didn't occur. In this sense the estimated magnitude was a failure. However, in a slightly longer time window, (20/2 – 1/5/2008), the equivalent seismic energy of a single seismic event of magnitude as Ms = 6.2R was released mainly through an unusual large number of EQs of Ms >=5R, more of less, magnitude. This can be explained by the fact that the seismicity of this time period consists part of an aftershock sequence (after the main seismic event) when a lot of seismic events take place. In the case when there is no previous main seismic event, then the actual magnitude of a future main large EQ, in the majority of the cases (Thanassoulas 2007), conforms to the magnitude calculation made by the application of the Lithospheric Seismic Energy Flow Model Methodology.

As far as it concerns the Methoni seismogenic region, it can be said that after the EQ of the 14/2/2008, the stored seismic energy decreased only for a small amount which equals to dMs = 0.11R. The later indicates that the Methoni area is still at a state of a potential future seismic threat. After the EQ of the 20/2/2008 (Ms = 6.5R) the regional Methoni seismogenic area remained "unlocked" for a time period of about 30 days. Then it was "locked" again. As a result, by taking into consideration this "unlocked" period, the Methoni seismogenic area became, in a very short time period, capable of generating a future EQ of Ms = 6.23R.

When large EQs will take place, in the future, in this area is generally unknown, as it is in every other known seismogenic area. However, what is made clear from this analysis and from other similar studies of preseismic electrical signals (Thanassoulas, 2007) is the fact (in most cases) that those large EQs, generally, a short time period, before their occurrence, <u>do generate preseismic electrical signals. Therefore, these EQs signal their presence in the near future, their location and their magnitude, quite accurately.</u>

**REFERENCES**

Thanassoulas, C., 2007. Short-term Earthquake Prediction, H. Dounias & Co, Athens, Greece. ISBN No: 978-960-930268-5

Thanassoulas, C., 2008. The electric field of the Earth after the occurrence of the February 14th, 2008, Ms = 6.7R EQ in Greece. Its implications towards the prediction of a probable future large EQ. arXiv.org:0802.3752 [pdf]

URL: www.earthquakeprediction.gr